\documentclass[12pt]{article}
\usepackage{graphicx}
\usepackage{amsfonts}
\usepackage{amssymb}
\usepackage{mathrsfs}
\usepackage{amsmath}
\usepackage[unicode]{hyperref}
\usepackage[numbers,sort&compress]{natbib}
\usepackage{color}

\begin{document}
\renewcommand{\thefootnote}{\fnsymbol{footnote}}
\begin{titlepage}

\vspace{10mm}
\begin{center}
{\Large\bf Self-regular black holes quantized by means of an analogue to hydrogen atoms}
\vspace{8mm}

{\large Chang Liu${}^{1,}$\footnote{E-mail address: liuchang13@mail.nankai.edu.cn},  Yan-Gang Miao${}^{1,2,3,}$\footnote{E-mail address: miaoyg@nankai.edu.cn (Corresponding author)}, Yu-Mei Wu${}^{1,}$\footnote{E-mail address: wuym@mail.nankai.edu.cn},  and Yu-Hao Zhang${}^{1,}$\footnote{E-mail address: yhzhang1994@gmail.com}

\vspace{6mm}
${}^{1}${\normalsize \em School of Physics, Nankai University, Tianjin 300071, China}

\vspace{3mm}
${}^{2}${\normalsize \em State Key Laboratory of Theoretical Physics, Institute of Theoretical Physics, \\Chinese Academy of Sciences, P.O. Box 2735, Beijing 100190, China}

\vspace{3mm}
${}^{3}${\normalsize \em CERN, PH-TH Division,
1211 Geneva 23, Switzerland}
}

\end{center}

\vspace{4mm}
\centerline{{\bf{Abstract}}}
\vspace{4mm}
We suggest a quantum black hole model that is based on an analogue to  hydrogen atoms.
A self-regular Schwarzschild-AdS black hole is investigated, where the mass density of the extreme black hole is given by the probability density of the ground state of hydrogen atoms and the mass densities of non-extreme black holes are given by the probability densities of excited states with no angular momenta. Such an analogue is inclined to adopt quantization of black hole horizons. In this way, the total mass of black holes is quantized. Furthermore, the quantum hoop conjecture and the Correspondence Principle are discussed.

\vskip 10pt
\noindent
{\bf PACS Number(s)}: 04.60.Bc, 04.70.Dy


\end{titlepage}
\newpage
\renewcommand{\thefootnote}{\arabic{footnote}}
\setcounter{footnote}{0}
\setcounter{page}{2}

 \section{Introduction}

It has been desirable that a non-perturbative quantum gravity theory should have no ultraviolet (UV) divergences~\cite{lnc2,lnc4,prd3,prd22}. To be consistent with this feature, the self-complete gravity theory~\cite{10053497,10101415} has been put forward to give a short distance cutoff that naturally avoids the UV divergence. Normally, the probe energy must be higher and higher when the exploration of microscope gets deeper and deeper. However, the production of micro  black holes provides a possible way to circumvent such an endless procedure. The UV self-completeness renders an intriguing property that micro black holes would be produced if elementary particle collisions with the Planckian energy scale could satisfy the so-called {\em quantum hoop conjecture}~\cite{s7}. When the energy goes higher, one cannot probe shorter distances but produce greater black holes with bigger horizons, where such objects are called classicalons~\cite{10053497,10101415,11035963,12052540,14097405}.
This means that the horizon of micro black holes gives a threshold or a natural minimal length
that might be probed experimentally.
On the other hand, if the minimal length implies the Planck length, the corresponding energy scale is far from being reached by the present and even foreseeing colliders. Nonetheless, if the effect of large extra dimensions could be considered, the micro black holes with the TeV scale would probably be produced at the LHC or its next generation in a not far future.
In the recent decade or so, there has been much progress on this issue, both in theory~\cite{0201228,0206060,0301073,0505181,0611224,08031287,08064647,09040230,10031798,
11105642,11012760,11115830,12021686,11105332,13097741,12100015,13022640} and in experiment~\cite{10123375,12026396,13084075,13112006,14054254}.

The idea of the self-completeness mentioned above has  been realized by Nicolini, Smailagic, and Spallucci~\cite{0510112} when  the noncommutative geometry~\cite{ncgeo} is introduced into the ordinary Schwarzschild black hole. It is assumed that the noncommutativity of spacetime would be an intrinsic rather than a super-imposed property of manifold, so that one should modify the distribution of matter and consequently the modified distribution of matter naturally reflects the basic characteristic of noncommutativity in manifold~\cite{14101706}. That is to say, energy-momentum tensors are modifed in terms of smeared matter distributions in the right hand side of Einstein's field equations, while no changes are made in the left hand side. By inserting the condition of energy conservation, $\nabla_\mu T^{\mu\nu}=0$, into a Schwarzschild-like solution, $g_{00}g_{rr}=-1$, a mass-smeared spherically symmetric black hole solution of the modified Einstein equations with a Gaussian mass density~\cite{0510112} can be obtained. The self-regular solution has no singularity at the origin and it naturally contains a minimal length that originates from the horizon of an extreme black hole. If the point-like mass distribution with the Dirac $\delta$-function density is taken, one obtains the ordinary Schwarzschild-AdS solution by solving the modified Einstein equations. For more details about this self-regular model of black holes, see refs.~\cite{0606051,0612035,08013519,08071939}.

Based on such a modification of mass distribution mentioned above, a new self-regular quantum black hole proposal~\cite{150301681} has been put forward by making an analogy between a self-regular black hole and a harmonic oscillator.
As the Gaussian mass density is proportional to the probability density of the ground state of a harmonic oscillator, the self-regular Schwarzschild black hole is regarded as a quantum harmonic oscillator. As a result, the total mass of the extreme self-regular Schwarzschild black hole is associated with the zero-point energy of a harmonic oscillator. In addition, the specific mass densities\footnote{Such mass densities are not set to be the probability densities of the excited states of a harmonic oscillator in order to avoid the appearance of multi-horizon geometries for the non-extreme black holes, which leads to the proposal being Bohr-like quantization, as explained in ref.~\cite{150301681}.}  with no multi-horizon geometries are chosen for non-extreme black holes. As it is assumed that the non-extreme black holes correspond to the excited states of a harmonic oscillator, the total masses of non-extreme black holes are thus associated with the energy eigenvalues of the excited states of a harmonic oscillator. Moreover, the quantum hoop conjecture and Correspondence Principle related to the analogy with a harmonic oscillator are found to be satisfied. The proposal briefly summarized above is named~\cite{150301681} as the Bohr-like quantization of the Schwarzschild black hole.

Inspired by the interesting Bohr-like quantization~\cite{150301681}, we propose in the present paper a so-called Schr\"odinger-like quantization for the self-regular Schwarzschild-AdS black hole. The meaning relies on our choice of black hole mass densities that depends on solutions of the Schr\"odinger equation not only for an extreme black hole but also for a non-extreme one. Although such a choice for a non-extreme black hole leads normally to the appearance of multi-horizon geometries, it is well-known that  an extreme  black hole, sometimes also named as a remnant of black holes at the final stage of the Hawking radiation, is quite different from a non-extreme one and therefore it seems to be far-fetched to require both of them mono-horizontal. The merit of our choice is that we provide a unified source of black hole mass densities for both the extreme and non-extreme cases, which makes our proposal succincter. Besides the formulations of mass densities, the other indispensable ingredient in the Bohr-like or our proposal is a special model of quantum mechanics that will be used to make an analogy with the black hole we are trying to quantize. Instead of a harmonic oscillator associated with the Bohr-like quantization~\cite{150301681}, we take a hydrogen atom as our specific model.
The reason of our choice comes from the recent works by Corda~\cite{12107747,13041899,150300565,150303474} and Bekenstein~\cite{150503253}, where the radiation spectrum of black holes is interpreted\footnote{We would like to point out that these works just established the analogy between the Hawking radiation spectra and the energy levels of a hydrogen atom, i.e. the quantization of the radiation spectra, in a semi-classical approach. However,  the quantization of the black hole itself was not touched, which leaves the task to the present paper.} to be similar to that of a hydrogen atom. That is, these works imply that there is a deep internal relationship between black holes and hydrogen atoms. Consequently, based on the Bohr-like quantization and the recent works by Corda and Bekenstein, we propose our scenario for quantization of the self-regular Schwarzschild-AdS black hole: the first step is to make the analogy between this black hole and the hydrogen atom, and then the second step is to choose the probability densities of states of hydrogen atoms to be the mass densities not only for an extreme black hole but also for a non-extreme one.

The arrangement of this paper is as follows. In the next section, we start from the metric of the self-regular Schwarzschild-AdS black hole, where the original total mass of a black hole has been replaced by a mass distribution. In this way, the noncommutativity of spacetime is introduced~\cite{0510112} into the Schwarzschild-AdS black hole and thus the curvature singularity at the origin is canceled. Further, the probability densities of the ground state and excited states of hydrogen atoms are chosen to be the mass densities of the extreme and non-extreme self-regular Schwarzschild-AdS black holes, where the ground state of hydrogen atoms corresponds to the extreme black hole and the excited states correspond to the non-extreme ones, which realizes the analogy between the self-regular Schwarzschild-AdS black hole and the hydrogen atom. Then, we analyze the mass quantization of the self-regular Schwarzschild-AdS black hole in section 3 through quantization of horizons. Such an analysis depends on the mean radius of hydrogen atoms, which is consistent with our specific analogy.  Moreover, the quantum hoop conjecture and the Correspondence Principle related to such an analogy are discussed. Finally, section 4 is devoted to a brief conclusion.

\section{Analogy between self-regular black holes and hydrogen atoms}

The metric of the static and spherically symmetric self-regular Schwarzschild-AdS black hole, where the noncommutativity of spacetime has been considered, takes the form,
\begin{equation}
ds^2=-\left(1-\dfrac{2\mathcal {M}(r)}{r}+\frac{r^2}{b^2}\right)dt^2+\left(1-\dfrac{2\mathcal {M}(r)}{r}+\frac{r^2}{b^2}\right)^{-1}dr^2+r^2\left(d{\theta}^2+{\sin}^2\theta d{\phi}^2\right), \label{metric}
\end{equation}
where the parameter $b$ is the radius of the AdS background spacetime. This metric is the so-called self-regular or noncommutative geometry inspired formulation of the Schwarzschild-AdS black hole with no metric and curvature singularities at the origin~\cite{0510112,0606051,0612035,08013519,08071939}.
The characteristic of this kind of black holes is that the mass distribution,
\begin{equation}
\mathcal {M}(r)=\int^r_0 \rho(r) 4\pi r^2 dr, \label{md}
\end{equation}
replaces the total mass, $M=\int^{\infty}_0 \rho(r) 4\pi r^2 dr$, in the metric. We shall see that the mass density $\rho(r)$ of black holes is related to a noncommutative parameter or a minimal length.

We emphasize that the metric solution (eq.~(\ref{metric})), as was shown in ref.~\cite{0510112}, is associated with the following modified energy-momentum tensor,
\begin{equation}
T^\mu{}_\nu=p_\bot \delta^\mu{}_\nu+(p_\bot+\rho)(u^\mu u_\nu-l^\mu l_\nu),
\end{equation}
where $u^\mu=\sqrt{g_{rr}}\delta^\mu_0$, $l^\mu=\frac{1}{\sqrt{g_{rr}}}\delta^\mu_r$, and $p_\bot=-\rho-\frac{r}{2}\frac{d\rho}{dr}$. Note that the appearance of the extra term $l^\mu l_\nu$ implies that the modified energy-momentum tensor describes a kind of anisotropic fluid rather than the perfect fluid. As a special case, when the point-like matter is taken, i.e. $\rho(r)=\frac{M}{2\pi r^2}\delta(r)$, one can retrieve the ordinary Schwarzschild-AdS solution by solving the modified Einstein equations rather than the Einstein equations.

\subsection{Analogy between the extreme black hole and the ground state}

According to our proposal, we take the probability density of the ground state of a hydrogen atom, $|\Psi_{100}|^2=\frac{1}{\pi  a_0^3} \exp({-\frac{2 r}{a_0}})$,
where $a_0$ is the Bohr radius, as the mass density for the extreme black hole,
\begin{equation}
\rho_1(r) =\frac{ M_1}{\pi  a^3}\exp\left({-\frac{2 r}{a}}\right),\label{massdensity1}
\end{equation}
where $M_1$ is the total mass of the extreme black hole and $a$ is a parameter that will be seen to be associated with the horizon radius of the extreme black hole, i.e. the minimal length in our model. We notice that the specific analogy between $|\Psi_{100}|^2$ and $\rho_1(r)$ is $a_0 \sim a$.

\begin{figure}[!ht]
\centering
\includegraphics[width=0.7\textwidth]{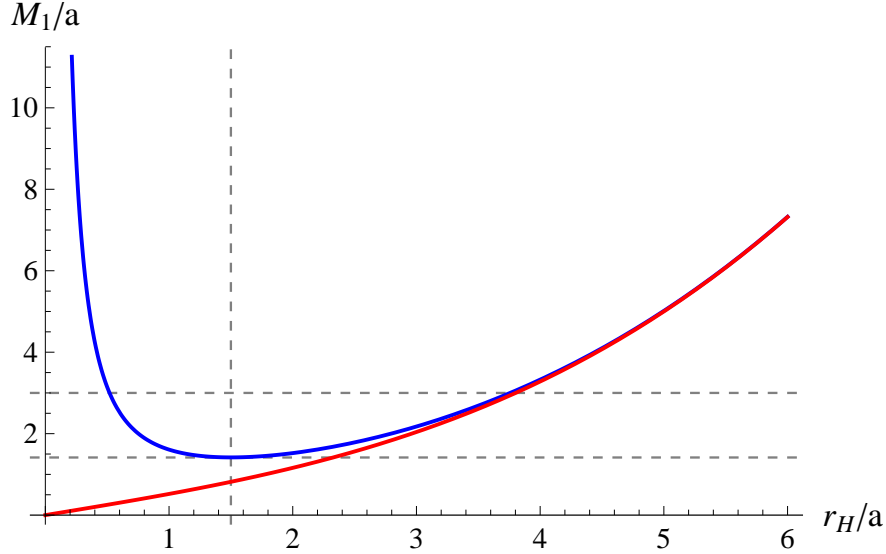}
\caption{The blue curve corresponds to the relation eq.~\eqref{mr} that gives the relation between the mass and the horizon of the extreme self-regular Schwarzschild-AdS black hole, and the red curve corresponds to the usual relation associated with  the ordinary Schwarzschild-AdS black hole. When the horizon radius grows up, the two curves gradually approach, which means that the effect of noncommutativity mainly exists in the near extremal horizon. Here we set $b=5a$, which satisfies the hoop conjecture, see below for the details.}
\label{m}
\end{figure}

Substituting eq.~(\ref{massdensity1}) into eq.~(\ref{md}), we obtain the mass distribution of the extreme black hole,
\begin{equation}
\mathcal {M}_1(r)=M_1\left[1-\left(1+\frac{2r}{a}+\frac{2 r^2}{a^2}\right)\exp\left({-\frac{2 r}{a}}\right)\right].\label{md1}
\end{equation}
One can see from eqs.~(\ref{metric}) and (\ref{md1}) that the metric singularity at $r=0$ has been canceled, which is consistent with the nonlocal gravity \cite{12022102}.
In addition, from $g_{00}=0$ we deduce the relation between the total mass $M_1$ and the horizon radius $r_{\rm H}$ as follows,
\begin{equation}
M_1=\frac{r_{\rm H}}{2} \left(1+\frac{r_{\rm H}^2}{b^2}\right)\left[1-\left(1+\frac{2 r_{\rm H}}{a}+\frac{2 r_{\rm H}^2}{a^2}\right)\exp\left({-\frac{2 r_{\rm H}}{a}}\right) \right]^{-1},\label{mr}
\end{equation}
which is plotted in Figure \ref{m}.

We can observe in Figure \ref{m} that there are two horizon radii in general, but for the extreme case where the mass takes the minimal value $M_1^{\rm min}$, there is only one horizon radius,  the extremal horizon radius $r_{{\rm H}_1}$. 
As $r_{{\rm H}_1}$ implies the minimal length, no horizon radius can approach zero. This is the characteristic of the self-regular black hole.

\begin{figure}[!ht]
\centering
\includegraphics[width=0.7\textwidth]{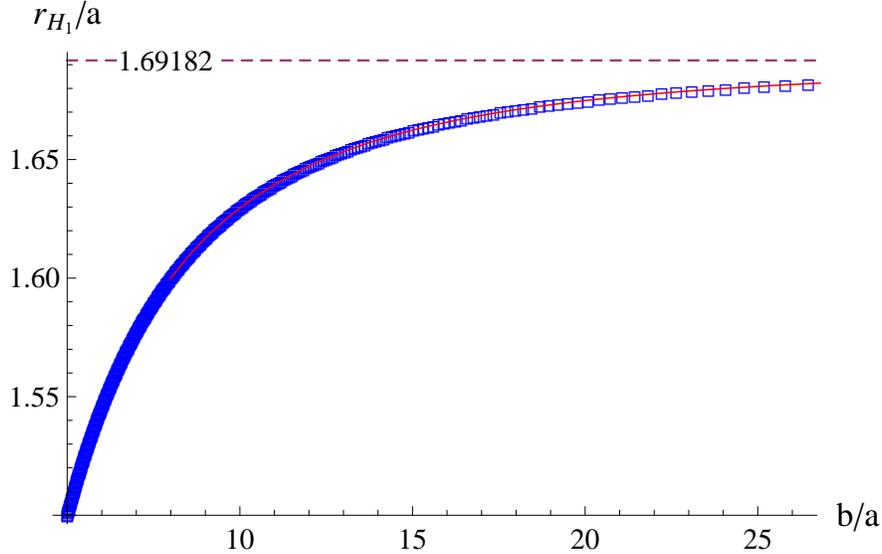}
\caption{The numerical points of eq.~(\ref{m00}) are plotted in blue color, and the fitting curve of eq.~(\ref{r0}) is plotted in red color. The relative error is less than $4\times10^{-5}\%$.}
\label{fit}
\end{figure}

By requiring ${\partial M_1}/{\partial r_{\rm H}}=0$, we find that the extremal horizon radius $r_{{\rm H}_1}$ that corresponds to $M_1^{\rm min}$ satisfies the following equation,
\begin{eqnarray}
& &\frac{4 r_{{\rm H}_1}^5}{a^5}+\frac{6 r_{{\rm H}_1}^4}{a^4}+\frac{6 r_{{\rm H}_1}^3}{a^3}+\frac{3 r_{{\rm H}_1}^2}{a^2}\left(1-\exp\left(\frac{2 r_{{\rm H}_1}}{a}\right)\right) \nonumber \\
& &
+\frac{b^2 }{a^2}\left[\frac{4 r_{{\rm H}_1}^3}{a^3}+\frac{2 r_{{\rm H}_1}^2}{a^2}+\frac{2 r_{{\rm H}_1}}{a}+\left(1- \exp\left(\frac{r_{{\rm H}_1}}{a}\right)\right)\right]=0, \label{m00}
\end{eqnarray}
where $r_{{\rm H}_1}$ can be regarded as the minimal length in our model.  
As eq.~(\ref{m00}) is a transcendental equation, one cannot solve it analytically. Therefore, we make a numerical fitting in terms of the rational fractional function,
\begin{equation}
\frac{r_{{\rm H}_1}}{a}=\frac{1.692 \left(\frac{b}{a}\right)^3+2.766 \left(\frac{b}{a}\right)^2+20.03 \frac{b}{a}-7.562}{\left(\frac{b}{a}\right)^3+1.635 \left(\frac{b}{a}\right)^2+15.94 \frac{b}{a}+3.198}. \label{r0}
\end{equation}
We plot eq.~(\ref{m00}) and its numerical fitting eq.~(\ref{r0}) for different ratios $b/a$ in Figure \ref{fit} from which we can see that the relative error is very small.

We now analyze the $b$-parameter dependence of the horizon radius of the extreme black hole.

(i) If $b \gg a$, which means an asymptotic Minkowski background, we compute from eqs.~(\ref{m00}) and (\ref{mr}) the extremal horizon  radius $r_{{\rm H}_1}\approx1.69182a$ and its corresponding minimal mass  $M_1^{\rm min}\approx1.28735a$.

(ii) In order to ensure the formation of a black hole, the hoop conjecture should be considered, that is, the mean radius of a black hole related to some mass distribution should not be larger than the horizon radius of the relevant extreme black hole.
The mean radius for the mass density eq.~(\ref{massdensity1}) reads
\begin{equation}
{\bar r}=\int_{0}^{\infty} r \rho_1(r)4\pi r^2 dr=\frac{3}{2} a.\label{meanr}
\end{equation}
Thus, the hoop conjecture requires $r_{{\rm H}_1} \geq \frac{3}{2} a$, whose lower bound gives the corresponding minimal mass $M_1^{\rm min}\approx1.41728a$. When we consider eq.~(\ref{m00}) or eq.~(\ref{r0}), the loop conjecture also implies the inequality of the ratio $b/a$, that is, $b/a \geq 4.99822$.

As a result, when the $b$-parameter of the AdS background spacetime meets the hoop conjecture, i.e. $4.99822 \leq b/a < \infty$, the horizon radius of the extreme black hole takes the following range,
\begin{equation}
\frac{3}{2}  \leq \frac{r_{{\rm H}_1}}{a} < 1.69182,\label{rhrange}
\end{equation}
and then the corresponding mass of the extreme black hole is constrained in the range,\footnote{We notice that for the extreme black hole a small radius corresponds to a large mass, and vice versa. The reason is that a small radius corresponds to a small $b$-parameter which gives rise to the large pressure $P$, $P \propto 1/b^2$, thus an extreme black hole with a small horizon radius has a high mass density and naturally it is heavier than an extreme black hole with a large horizon radius.} $1.28735a < M_1^{\rm min} \leq 1.41728a$.
This implies that the AdS radius cannot be too small, or the curvature of the background spacetime cannot be too large. Moreover, although $b$ has a wide range, ${r_{{\rm H}_1}}/{a}$ has a narrow one. That is, ${r_{{\rm H}_1}}/{a}$ correlates weakly with $b$. As mentioned above, we may take $l_0  = \frac{3}{2} a$ as the minimal length which appears naturally from the horizon radius of the extreme black hole.

\subsection{Analogy between non-extreme black holes and excited states}

In accordance with our proposal mentioned in the above section, we take the probability densities of excited states of hydrogen atoms as the mass densities of non-extreme black holes. As our black hole is non-rotational, we choose the probability densities of excited states with no angular momenta $|\Psi_{n00}|^2$ to be the desired mass densities $\rho_n(r)$,
\begin{equation}
\rho_n(r)= \frac{M_n}{\pi n^5 a^3 }  \left(\sum _{k=0}^{n-1} \frac{n!}{(n-k-1)!(k+1)!k!} \left(-\frac{2 r}{na}\right)^k\right)^2\exp\left({-\frac{2 r}{na}}\right), \label{rhon}
\end{equation}
where $n$ is a positive integer, and $M_n$ is the total mass of the non-extreme black hole related to the $n$-th energy level of excited states with no angular momenta. This formula includes the ground state to be the special case of $n=1$.
Although such a choice of the mass densities for non-extreme black holes will lead to multi-horizon solutions, there is no evidence that both the extreme and non-extreme black holes would be mono-horizontal, and this choice provides a unified source of black hole mass distributions for both the extreme and the non-extreme cases.

Substituting eq.~(\ref{rhon}) into eq.~(\ref{md}), we compute the mass distributions of non-extreme black holes with $n \geq 2$,
\begin{eqnarray}
\mathcal {M}_2(r) &=& M_2\left[1-\left(1+\frac{r}{a}+\frac{r^2}{2 a^2}+\frac{r^4}{8 a^4}\right)\exp\left({-\frac{r}{a}}\right)\right],\\
\mathcal {M}_3(r) &=& M_3\left[1-\left(1+\frac{2 r}{3 a}+\frac{2 r^2}{9 a^2}+\frac{4 r^4}{81 a^4}-\frac{8 r^5}{729 a^5}+\frac{8 r^6}{6561 a^6}\right)\exp\left({-\frac{2r}{3a}}\right)\right],\\
\mathcal {M}_4(r) &=& M_4\left[1-\left(1+\frac{r}{2 a}+\frac{r^2}{8 a^2}+\frac{3 r^4}{128 a^4}-\frac{r^5}{128 a^5}+\frac{13 r^6}{9216 a^6}-\frac{r^7}{9216 a^7}\right.\right. \nonumber \\
& &\;\;\;\;\;\;\;\;\; \left.\left. +\frac{r^8}{294912 a^8}\right)\exp\left({-\frac{r}{2a}}\right)\right],\\
& \vdots & \nonumber
\end{eqnarray}
Setting the largest real root $r_{{\rm H}_n}$ of $g_{00}=0$ be the horizon radius of the $n$-th non-extreme black hole, we express the total mass $M_n$ in terms of the corresponding horizon radius $r_{{\rm H}_n}$,
\begin{eqnarray}
{M}_2 &=&  \frac{r_{{\rm H}_2}}{2} \left(1+\frac{r_{{\rm H}_2}^2}{b^2}\right)\left[1-\left(1+\frac{r_{{\rm H}_2}}{a}+\frac{r_{{\rm H}_2}^2}{2 a^2}+\frac{r_{{\rm H}_2}^4}{8 a^4}\right)\exp\left({-\frac{r_{{\rm H}_2}}{a}}\right)\right]^{-1},\label{m2}\\
{M}_3 &=& \frac{r_{{\rm H}_3}}{2} \left(1+\frac{r_{{\rm H}_3}^2}{b^2}\right)\left[1-\left(1+\frac{2 r_{{\rm H}_3}}{3 a}+\frac{2 r_{{\rm H}_3}^2}{9 a^2}+\frac{4 r_{{\rm H}_3}^4}{81 a^4}-\frac{8 r_{{\rm H}_3}^5}{729 a^5}\right.\right. \nonumber \\
& &\;\;\;\;\;\;\;\;\;\;\;\;\;\;\;\;\;\;\;\;\;\;\;\;\;\;\;\; \left.\left.+\frac{8 r_{{\rm H}_3}^6}{6561 a^6}\right)\exp\left(-\frac{2 r_{{\rm H}_3}}{3a}\right)\right]^{-1},\label{m3}\\
{M}_4 &=& \frac{r_{{\rm H}_4}}{2} \left(1+\frac{r_{{\rm H}_4}^2}{b^2}\right)\left[1-\left(1+\frac{r_{{\rm H}_4}}{2 a}+\frac{r_{{\rm H}_4}^2}{8 a^2}+\frac{3 r_{{\rm H}_4}^4}{128 a^4}-\frac{r_{{\rm H}_4}^5}{128 a^5}+\frac{13 r_{{\rm H}_4}^6}{9216 a^6}\right.\right. \nonumber \\
& &\;\;\;\;\;\;\;\;\;\;\;\;\;\;\;\;\;\;\;\;\;\;\;\;\;\;\;\; \left.\left.-\frac{r_{{\rm H}_4}^7}{9216 a^7}+\frac{r_{{\rm H}_4}^8}{294912 a^8}\right)\exp\left(-\frac{r_{{\rm H}_4}}{2a}\right)\right]^{-1},\label{m4}\\
& \vdots & \nonumber
\end{eqnarray}
It is now ready for us to quantize $M_n$ by means of quantization of $r_{{\rm H}_n}$.

\section{Quantization of extreme and non-extreme black holes}

In ref.~\cite{150301681} the mass of black holes is quantized directly because the self-regular Schwarzschild black hole is regarded as the quantum harmonic oscillator. Here the situation is different. Our proposal, based on the works by Corda~\cite{12107747,13041899,150300565,150303474} and Bekenstein~\cite{150503253}, is the analogue of the self-regular Schwarzschild-AdS  black hole and the hydrogen atom with no angular momenta. Thus, we are inclined to adopt quantization of horizons. Specifically, for the hydrogen atom with no angular momenta its quantum mean radius reads $ \langle r  \rangle=\frac{3a_0}{2}n^2\propto n^2$. Because the mean radius of hydrogen atoms corresponds to the horizon radius of black holes,
the quantum horizon radius of the self-regular Schwarzschild-AdS  black hole is naturally assumed to be
\begin{equation}
r_{{\rm H}_n}=n^2 r_{{\rm H}_1}, \label{rhn}
\end{equation}
where $r_{{\rm H}_1}$ is, like $a_0$ in hydrogen atoms, the horizon radius of the extreme black hole. Substituting eq.~(\ref{rhn}) into eqs.~(\ref{mr}) and (\ref{m2})-(\ref{m4}), we obtain the quantized masses of the extreme and non-extreme black holes that are expressed in terms of the extremal horizon radius or the minimal length $r_{{\rm H}_1}$,
\begin{eqnarray}
M_1^{\rm min} &=& \frac{r_{{\rm H}_1}}{2} \left(1+\frac{r_{{\rm H}_1}^2}{b^2}\right)\left[1-\left(1+\frac{2 r_{{\rm H}_1}}{a}+\frac{2 r_{{\rm H}_1}^2}{a^2}\right)\exp\left({-\frac{2 r_{{\rm H}_1}}{a}}\right) \right]^{-1},\\
M_2^{\rm quan} &=& 2 r_{{\rm H}_1} \left(1+\frac{16 r_{{\rm H}_1}^2}{b^2}\right)\left[1-\left(1+\frac{4 r_{{\rm H}_1}}{a}+\frac{8 r_{{\rm H}_1}^2}{a^2} +\frac{32 r_{{\rm H}_1}^4}{a^4}\right)\exp\left(-\frac{4 r_{{\rm H}_1}}{a}\right)\right]^{-1},\\
M_3^{\rm quan} &=&\frac{9 r_{{\rm H}_1}}{2} \left(1+\frac{81 r_{{\rm H}_1}^2}{b^2}\right)\left[1-\left(1+\frac{6 r_{{\rm H}_1}}{a}+\frac{18 r_{{\rm H}_1}^2}{2 a^2}+\frac{324 r_{{\rm H}_1}^4}{a^4}-\frac{648 r_{{\rm H}_1}^5}{a^5} \right.\right.\nonumber \\
& & \;\;\;\;\;\;\;\;\;\;\;\;\;\;\;\;\;\;\;\;\;\;\;\;\;\;\;\;\;\;\;\;\;\;\;\;\;\;\;\;\; \left.\left.+\frac{648 r_{{\rm H}_1}^6}{a^6}\right)\exp\left(-\frac{6 r_{{\rm H}_1}}{a}\right)\right]^{-1}, \\
M_4^{\rm quan} &=& 8 r_{{\rm H}_1} \left(1+\frac{256 r_{{\rm H}_1}^2}{b^2}\right)\left[1-\left(1+\frac{8 r_{{\rm H}_1}}{a}+\frac{32 r_{{\rm H}_1}^2}{a^2}+\frac{1536 r_{{\rm H}_1}^4}{a^4}-\frac{8192 r_{{\rm H}_1}^5}{a^5}+\frac{212992 r_{{\rm H}_1}^6}{9 a^6} \right.\right.\nonumber \\
& & \;\;\;\;\;\;\;\;\;\;\;\;\;\;\;\;\;\;\;\;\;\;\;\;\;\;\;\;\;\;\;\;\;\;\;\;\;\;\;\;\;\;\; \left.\left.-\frac{262144 r_{{\rm H}_1}^7}{9 a^7}+\frac{131072 r_{{\rm H}_1}^8}{9 a^8}\right)\exp\left(-\frac{8 r_{{\rm H}_1}}{a}\right)\right]^{-1},\\
& \vdots &  \nonumber
\end{eqnarray}

\begin{figure}[!ht]
\centering
\includegraphics[width=0.8\textwidth]{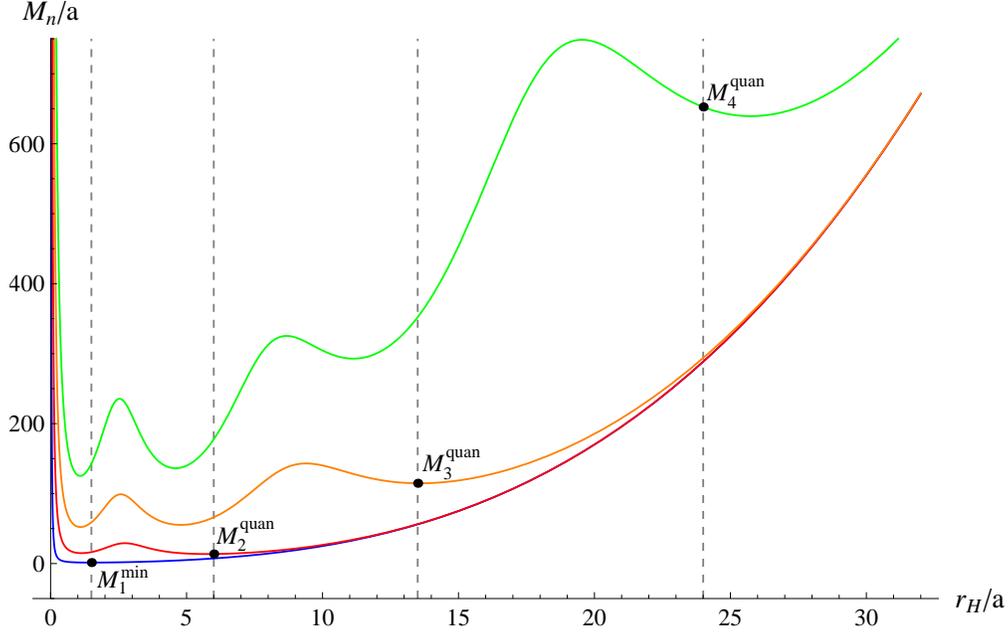}
\caption{Plots of the relations between black hole masses and their horizon radii. The blue, red, orange, and green curves correspond to the cases of $n=1,2,3,4$, respectively, where we set $b/a=5$ which has been verified to satisfy the hoop conjecture, see eq.~(\ref{meanr}).}
\label{mn}
\end{figure}

We plot eqs.~(\ref{mr}) and (\ref{m2})-(\ref{m4}) in Figure \ref{mn}, where the four black round points denote the quantized masses of the extreme black hole ($M_1^{\rm min}$) and the non-extreme black holes ($M_n^{\rm quan}, n=2,3,4$), respectively.

Now we turn to the discussion of the quantum hoop conjecture which has the following form~\cite{150301681},
\begin{equation}
 \langle n| r |n\rangle \leq \langle n| r_{{\rm H}_n} |n\rangle.\label{qhc}
\end{equation}
Considering eqs.~(\ref{rhon}) and (\ref{rhn}), we calculate
\begin{eqnarray}
 \langle n| r |n\rangle &=& \int_0^\infty r\, \frac{\rho_n(r)}{M_n} 4 \pi r^2  dr = \frac{3}{2}  a n^2,\\
\langle n| r_{{\rm H}_n} |n\rangle &=& \int_0^\infty r_{{\rm H}_n} \frac{\rho_n(r)}{M_n} 4 \pi r^2  dr =n^4r_{{\rm H}_1}.
\end{eqnarray}
As a result, the quantum hoop conjecture in our proposal reads
\begin{equation}
\frac{3}{2}  a \leq n^2 r_{{\rm H}_1}. \label{qhcrn}
\end{equation}
For the extreme black hole, i.e. the case $n=1$, the quantum hoop conjecture is satisfied because it reduces to eq.~(\ref{rhrange}). For the non-extreme black holes, i.e. the cases $n \geq 2$, the quantum hoop conjecture is obviously satisfied. This  means that our assumption of quantization, see eq.~(\ref{rhn}), coincides with the quantum hoop conjecture. That is to say, the black holes can be formed at the quantum level in our proposal.

As to the Correspondence Principle, it usually indicates a transition from quantum theory to classical theory. In quantum mechanics, there are two alternatives to realize such a transition. One is the limit of a large quantum number, and the other is the limit of $\hbar \rightarrow 0$. The latter alternative corresponds to the limit of $a \rightarrow 0$ in our proposal, which implies the fact that the minimal length ($l_0 = \frac{3}{2} a$) can be neglected for a black hole with a large scale. We can check that  when $a \rightarrow 0$, the mass densities (eq.~(\ref{rhon})) turn back to the $\delta (r)$-function density that describes, from the point of view of the modified Einstein equations, the ordinary Schwarzschild-AdS black hole without the effect of the minimal length.
Consequently, the Correspondence Principle is satisfied in our proposal of black hole quantization.

\section{Summary}

Based on the recent works by Corda~\cite{12107747,13041899,150300565,150303474} and Bekenstein~\cite{150503253}, the analogue of a self-regular Schwarzschild-AdS black hole and a hydrogen atom is assumed. Correspondingly, the quantization of horizons is utilized. In this way, the total mass of a self-regular Schwarzschild-AdS black hole is quantized. Moreover,
the quantum hoop conjecture and the Correspondence Principle are verified in our proposal.


\section*{Acknowledgments}
CL would like to thank X. Hao and L. Zhao of Nankai University for their helpful discussions. Y-GM would like to thank W. Lerche of PH-TH Division of CERN for kind hospitality. This work was supported in part by the National Natural
Science Foundation of China under grant No.11175090 and
by the Ministry of Education of China under grant No.20120031110027.
At last, the authors would like to thank the anonymous referees for their helpful comments that indeed improve this work greatly.

\end{document}